\documentstyle[11pt,blois,epsfig,rotating]{article}

\def\Journal#1#2#3#4{{#1} {\bf #2} (#3) #4}

\newcommand{\btab}{\begin{tabular}}
\newcommand{\etab}{\end{tabular}}
\newcommand{\beq}{\begin{equation}}
\newcommand{\eeq}{\end{equation}}
\newcommand{\beqn}{\begin{eqnarray}}
\newcommand{\eeqn}{\end{eqnarray}}
\newcommand{\be}{\begin{equation}}
\newcommand{\ee}{\end{equation}}
\newcommand{\bea}{\begin{array}}
\newcommand{\ena}{\end{array}}
\newcommand{\begitm}{\begin{itemize}}
\newcommand{\eitm}{\end{itemize}}

\def\epem{\mbox{$e^+e^-$}}

\def\gevcc{\mbox{GeV/$c^2$}}

\def\infb{\mbox{$\mbox{fb}^{-1}$}}

\def\epem{\mbox{$e^+e^-$}}

\newcommand{\AmS}{{\protect\the\textfont2
  A\kern-.1667em\lower.5ex\hbox{M}\kern-.125emS}}

\def\infb{\mbox{$\hbox{fb}^{-1}$}}

\def\epem{\mbox{$\hbox{e}^+\hbox{e}^-$}}
\def\mpmm{\mbox{$\mu^{+}\mu^{-}$}}
\def\tptm{\mbox{$\tau^+\tau^-$}}
\def\lplm{\mbox{$\ell^+\ell^-$}}

\def\ffbar{\mbox{$\hbox{f}\bar{\hbox{f}}$}}

\def\bbbar{\mbox{$\hbox{b}\bar{\hbox{b}}$}}

\def\suppeg{\hbox{\lower -.08cm \hbox{
{\hbox{$>$}}{\hbox{\kern -.30cm\lower .18cm \hbox{$\sim$}}}}}}
\def\infpeg{\hbox{\lower -.08cm \hbox{
{\hbox{$<$}}{\hbox{\kern -.50cm\lower .28cm \hbox{$\sim$}}}}}}

\def\suppeg{\hbox{\lower -.08cm \hbox{
{\hbox{$>$}}{\hbox{\kern -.26cm\lower .18cm \hbox{$\sim$}}}}}}
\def\infpeg{\hbox{\lower -.08cm \hbox{
{\hbox{$<$}}{\hbox{\kern -.30cm\lower .24cm \hbox{$\sim$}}}}}}

\def\PLB{{\em Phys. Lett.}  {\bf B}}

\def\PRD{{\em Phys. Rev.} {\bf D}}

\def\EPJC{{\em Eur. Phys. J.} {\bf C}}
\def\JPG{{\em J. Phys.} {\bf G}}

\def\be{\begin{equation}}
\def\ee{\end{equation}}
\def\bea{\begin{eqnarray}}
\def\eea{\end{eqnarray}}

\begin{document}

\begin{picture}(80,100)
\put(-40,-210){\epsfxsize220mm\epsfbox{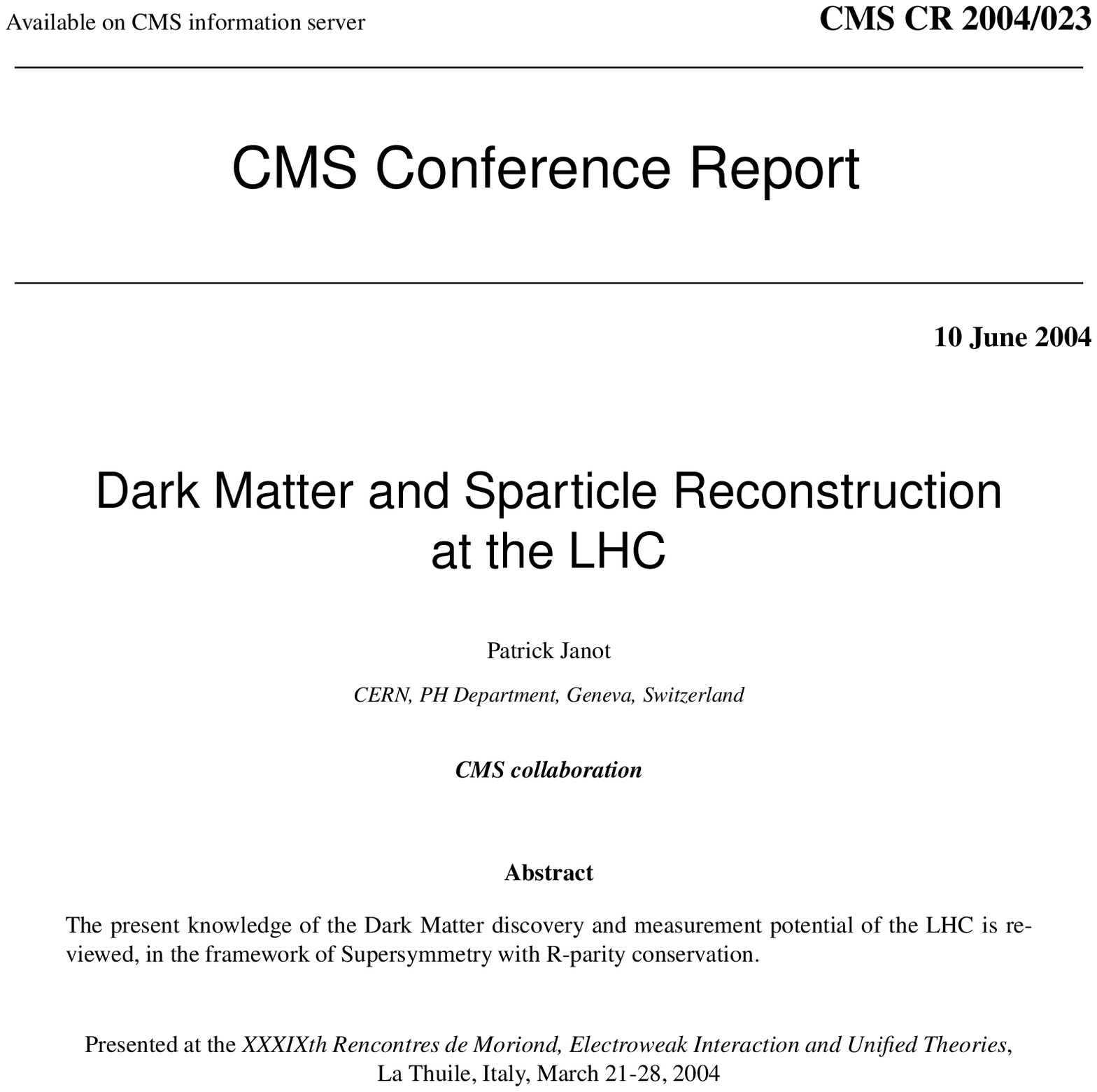}}
\end{picture}
\eject\null

\vspace*{4cm}
\title{DARK MATTER AND SPARTICLE RECONSTRUCTION AT THE LHC}

\author{Patrick JANOT}
\address{CERN, PH Department, CH-1211 Geneva 23, Switzerland}

\maketitle
\abstracts{The present knowledge of the Dark Matter discovery and 
measurement potential of the LHC is reviewed, in the framework of 
Supersymmetry with R-parity conservation.}

\section{Introduction}
Until now, the search for Dark Matter at accelerators (LEP, TeVatron) has 
been unsuccessful. Although unclear hints have been claimed for years by 
the DAMA experiment~\cite{dama} --- and are being contradicted by other 
experiments~\cite{fiorucci} --- the vast majority of firm measurements 
regarding Dark Matter has come from Cosmology. The accuracy of the 
Cosmic Microwave Background (CMB) observations has recently improved 
very significantly with the contribution of WMAP~\cite{wmap}. The latter 
confirmed many earlier and less accurate determinations of the Cold 
Dark Matter (CDM) density, $\Omega_{\rm m} h^2$, in the Universe, which 
is now constrained to lie in the 95\%\,C.L. interval
\begin{equation}
0.093 < \Omega_{\rm m} h^2 < 0.129, 
\label{eq:wmap}
\end{equation}
in units of the critical density $\rho_{\rm c}=1.86\,10^{-29}\,{\rm g/cm}^3$.

The most popular candidate for this Dark Matter, which complies with the 
electroweak precision measurements performed at LEP and SLC, appears in 
Supersymmetry (SUSY). With R-parity conservation, the lightest supersymmetric 
particle (LSP) is stable and its relic density may indeed lie in the above 
interval for suitable choices of SUSY parameters.

For the sake of definiteness and simplicity (and because most of the studies 
have yet been performed under these assumptions), the Minimal Supersymmetric 
extension of the Standard Model (MSSM) is chosen as the theoretical framework
in this writeup. 
The LSP, which has to be neutral and colourless to be a reliable candidate 
for CDM, is assumed to be the ligthest neutralino, $\chi^0_1$. 
The mSugra-inspired SUSY-breaking mechanism is assumed throughout. Minimal 
Supergravity is described by only four parameters: the universal scalar mass 
$m_0$, the universal gaugino mass $m_{1/2}$, the ratio of the two 
Higgs-doublet vacuum expectation values $\tan\beta$, and the sign of the 
Higgs mixing parameter $\mu$. To simplify the picture, a fifth parameter, 
the trilinear SUSY-breaking parameter $A_0$, was fixed to zero.

This report is organized as follows. In Section~\ref{sec:cosmo}, 
the constraints from the CMB measurements on the SUSY parameters 
are summarized and are then turned into a list of benchmark points 
(compatible with these measurements) for LHC to study. The LHC study 
of the most favourable of these benchmark points is presented in 
Section~\ref{sec:ptb}, in the simplified framework described above. 
More difficult situations, as well as the possible consequences of 
the simplifying theoretical assumptions are briefly discussed in 
Section~\ref{sec:difficult}.

\section{Dark Matter: Cosmology and Supersymmetry}
\label{sec:cosmo}
If the CDM density hinted at by the CMB measurements is
solely made of the lightest neutralino $\chi^0_1$, it is equal to the 
product of the relic LSP density, $n_{\rm LSP}$, and the LSP mass,
$m_{\rm LSP}$,
\begin{equation}
\rho_{\rm LSP} = \Omega_{\rm m} h^2 = n_{\rm LSP} \times m_{\rm LSP}.
\label{eq:rho}
\end{equation}
The measured $\Omega_{\rm m} h^2$ interval of Eq.~\ref{eq:wmap} 
therefore allows this product to be constrained. This, as is 
shown below, constrains in turn the LSP mass to be rather small. 

\subsection{The annihilation bulk}

\begin{figure}[ht]
\begin{picture}(160,35)
\put(0,-5){\epsfxsize50mm\epsfbox{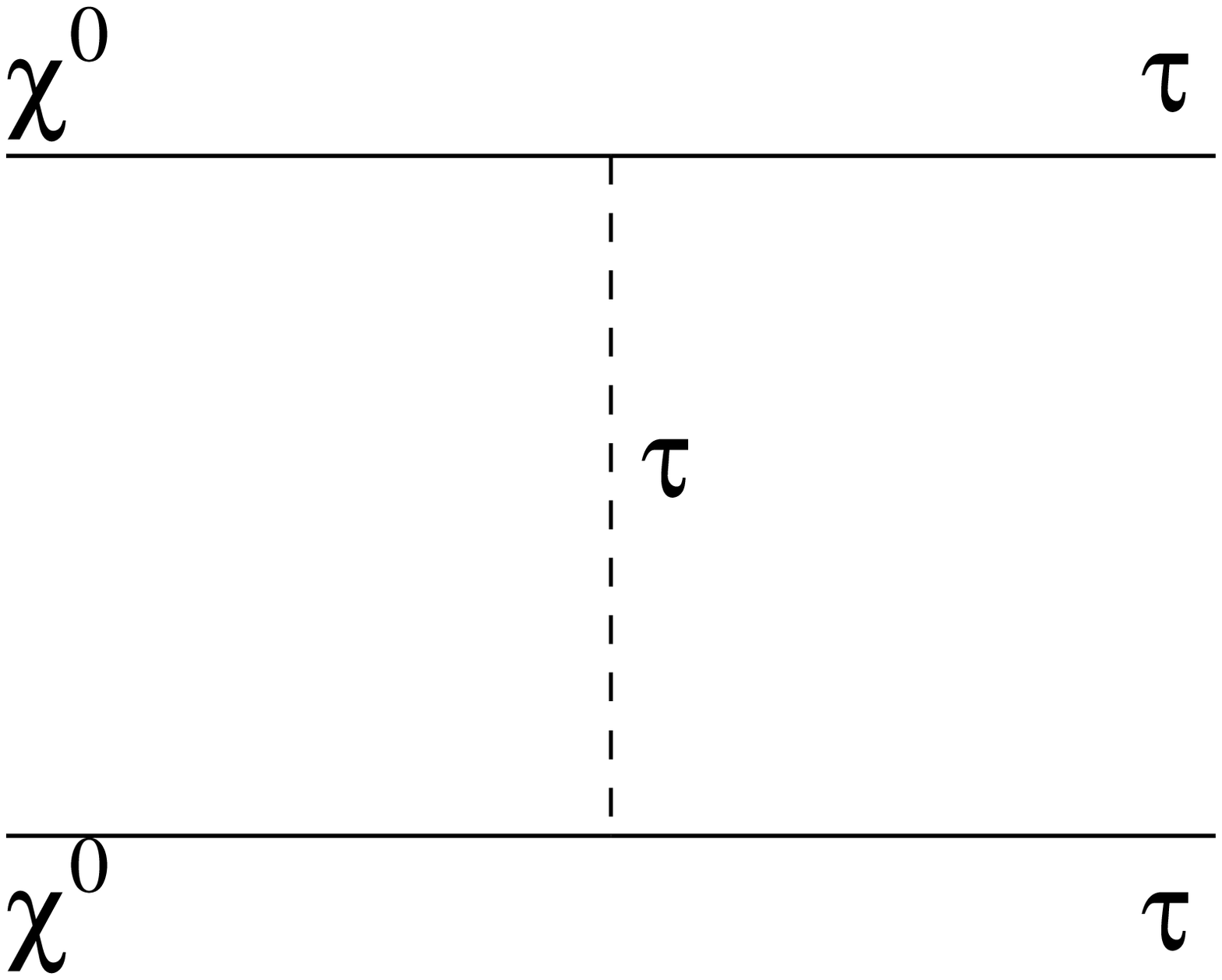}}
\put(55,-5){\epsfxsize50mm\epsfbox{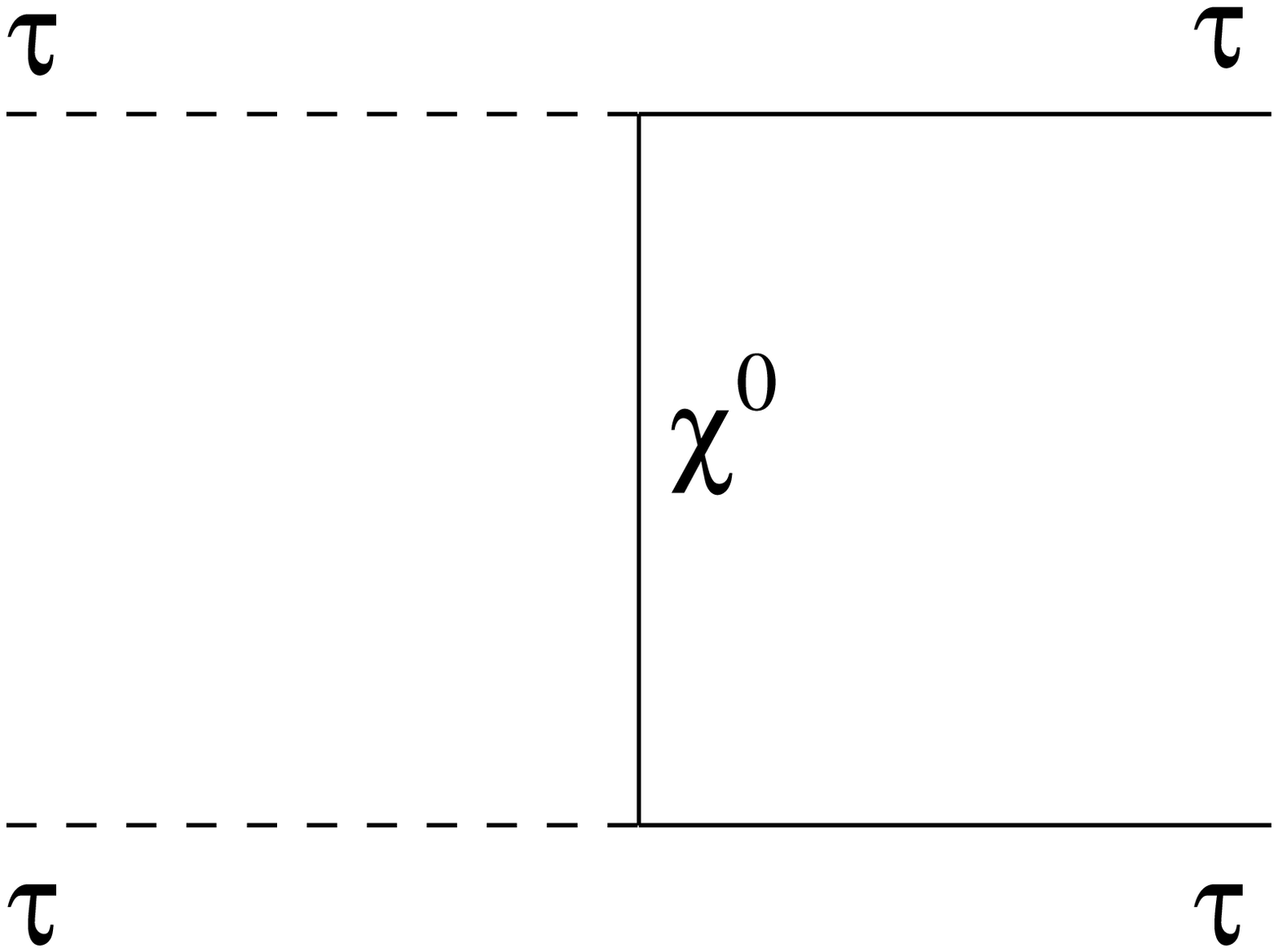}}
\put(110,-5){\epsfxsize50mm\epsfbox{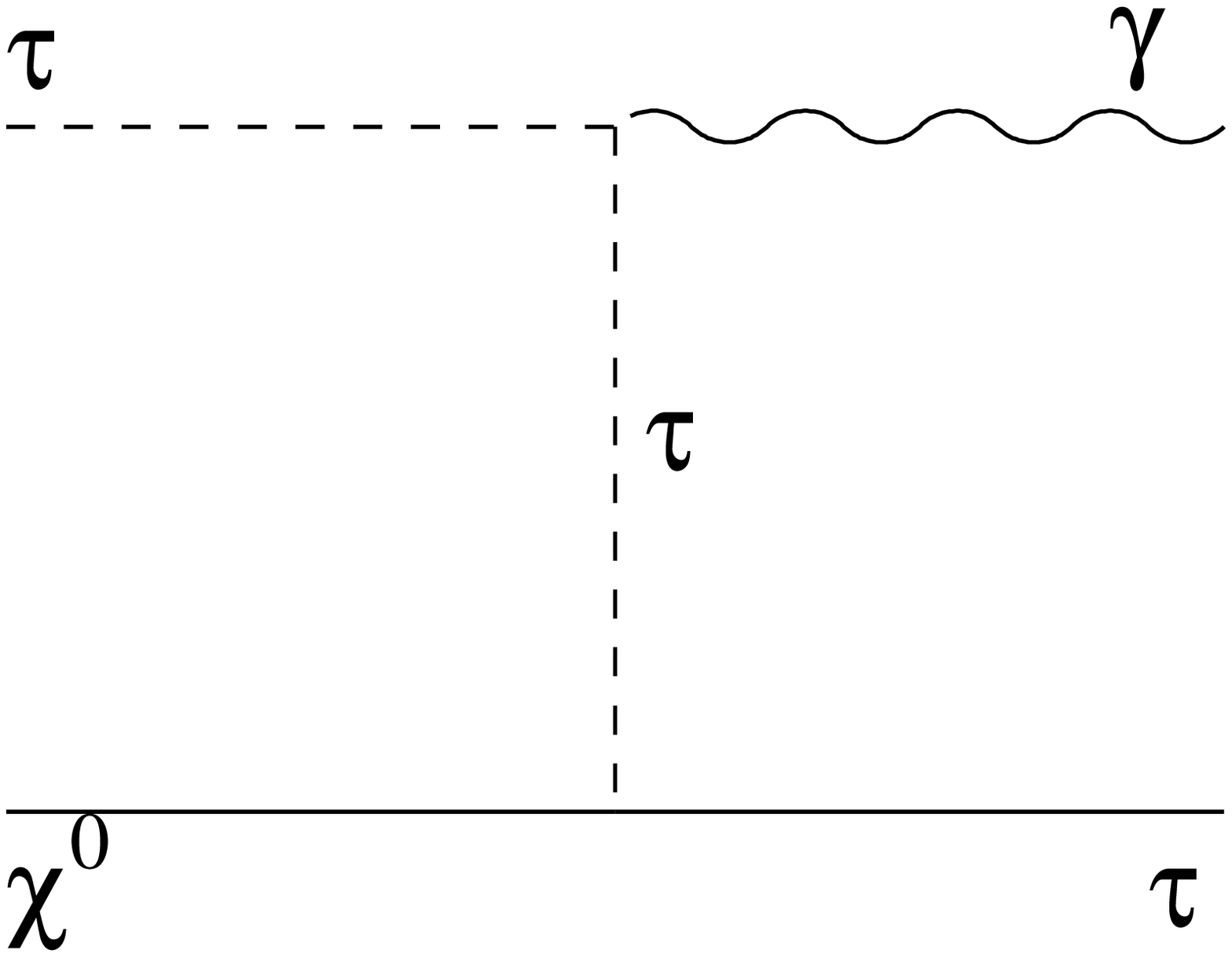}}
\put(25,22){$\sim$}
\put(59,6.7){$\sim$}
\put(59,34){$\sim$}
\put(114,34){$\sim$}
\put(135.5,22){$\sim$}
\put(22,0){(a)}
\put(77,0){(b)}
\put(132,0){(c)}
\end{picture}
\caption{(a) Dominant process for the LSP annihilation through a 
$t$-channel exchange of the NSLP, assumed here to be the lighter
$\tilde\tau$. (b) and (c) Dominant diagrams for $\tilde{\tau}$-$\chi^0_1$ 
co-annihilation in the quasi-mass-degenerate configuration}
\label{fig:coan}
\end{figure}

The relic LSP density is expected to be inversely proportional to 
the cross section for the LSP annihilation in the early universe. 
In general, the latter is expected to be dominated by the annihilation 
into fermion pairs, $\chi_1^0 \chi^0_1 
\to \ffbar$, through the $t$-channel exchange of the lightest sfermion 
$\tilde {\rm f}$, the next-to-lightest supersymmetric particle (NLSP). 
This annihilation proceeds according to the graph of Fig.~\ref{fig:coan}a,
with a cross section proportional to 
$m_\chi^2/(m_\chi^2+m_{\tilde {\rm f}}^2)^2$. According to Eq.~\ref{eq:rho}, 
the resulting CDM density is therefore proportional to 
$(m_\chi^2+m_{\tilde {\rm f}}^2)^2/m_\chi$, {\it i.e.}, about proportional
to $m_\chi^3$ if the NLSP and the LSP masses are of the same order. 
As a consequence, Eq.~\ref{eq:wmap} allows lower and upper limits
to be set on the LSP mass (or a combination of the LSP and the NLSP 
masses) or, equivalently, on a combination of $m_0$ and $m_{1/2}$. 

The upper limits are, for the bulk of the parameter sets (called the 
{\it annihilation bulk}), of the order of 150\,\gevcc\ 
for $m_0$ and 300\,\gevcc\ for $m_{1/2}$, with the consequence that 
all sparticles are expected to be light  --- with an upper limit on 
the LSP mass of the order of 150\,\gevcc\ --- hence to be produced 
copiously at the LHC.

\subsection{The co-annihilation tail}

This general picture is modified in a few corners of the parameter space, 
in which the total LSP-annihilation cross section can be made substantially  
larger than in the annihilation bulk. Such exotic configurations happen when 
the LSP and the NLSP, {\it i.e.}, the lighter $\tilde \tau$ in mSugra, are 
almost degenerate. Indeed, if the mass difference $m_{\tilde\tau}-m_{\chi^0_1}$
is smaller than the $\tau$ mass, no direct decay of the $\tilde\tau$ into 
$\tau\chi^0_1$ can take place. The $\tilde\tau$ disappears slowly via 
the annihilation diagram of Fig.~\ref{fig:coan}b, $\tilde\tau \tilde\tau 
\to \tau\tau$, and the co-annihilation process of  Fig.~\ref{fig:coan}c, 
$\tilde\tau \chi^0_1 \to \tau\gamma$ can proceed in parallel. The LSP 
relic density is reduced in turn, which allows larger values 
for the LSP mass and increases the upper limit on $m_{1/2}$, typically 
of 1\,TeV/$c^2$ or slightly more~\cite{coann}.

\subsection{The rapid annihilation funnels and the focus points}

\begin{figure}[ht]
\begin{picture}(160,70)
\put(0,0){\epsfxsize70mm\epsfbox{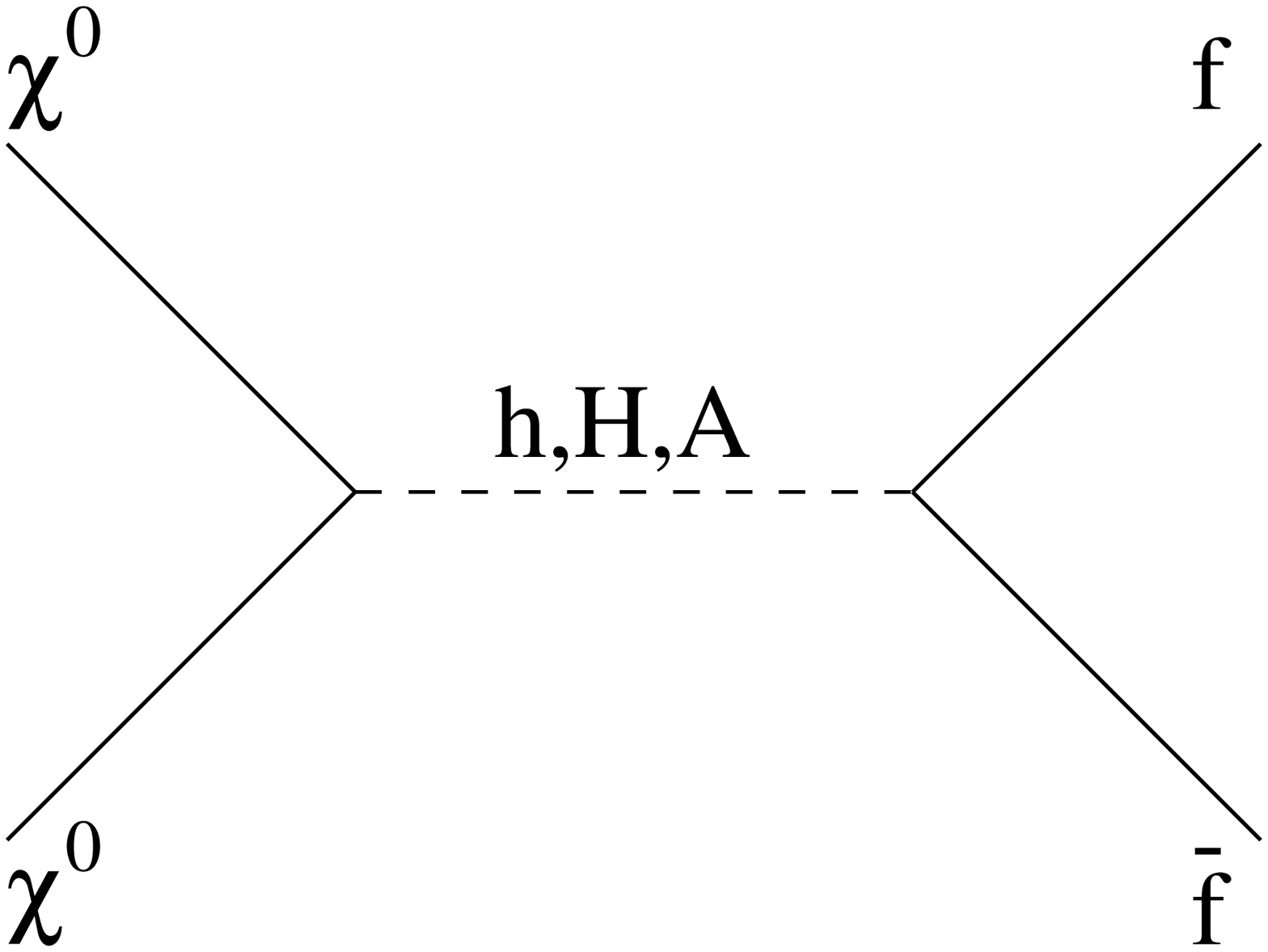}}
\put(80,-3){\epsfxsize80mm\epsfbox{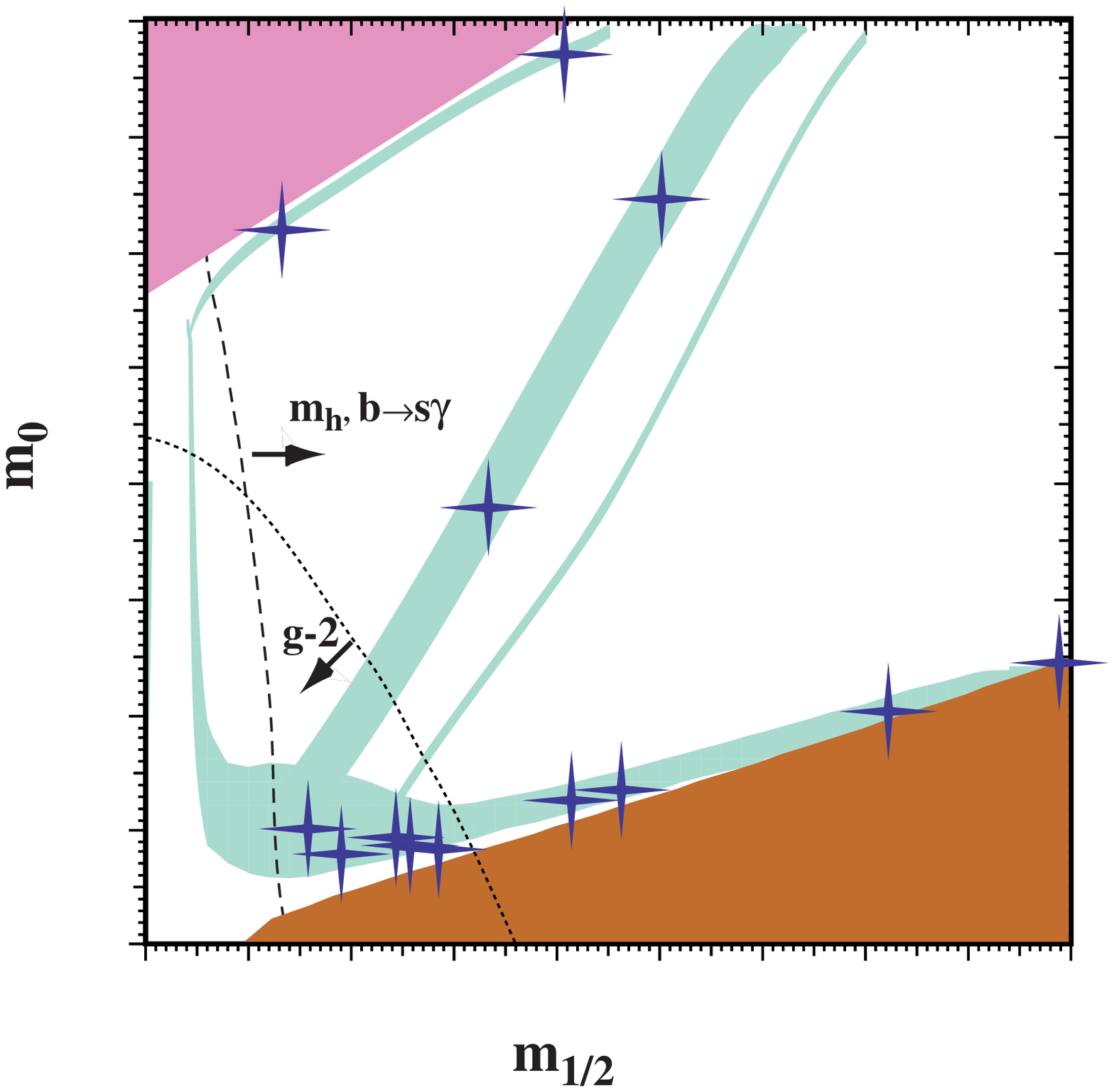}}
\put(31,55){(a)}
\put(145,60){(b)}
\put(98,62){\small Focus Points}
\put(121,52){\small Rapid}
\put(116,48){\small Annihilation}
\put(120,44){\small Funnels}
\put(120,11){\small Charged LSP (stau)}
\put(104,20){\small Bulk}
\put(120,20){\begin{rotate}{15}
{\small Co-annihilation tail}\end{rotate}}
\end{picture}
\caption{(a) Annihilation via Higgs boson exchange, relevant in the rapid
annihilation funnels (on-shell annihilation) and in the focus points 
(off-shell annihilation); (b) 
Qualititative overview of the location of the various regions in the $(m_{1/2},
m_0)$ plane, populated by a few benchmark points, taken from Ref.$^7$.}
\label{fig:rapid}
\end{figure}

Neutralino annihilation may also proceed via the graph of 
Fig.~\ref{fig:rapid} through an exchange of Higgs boson(s) in the $s$ 
channel. This diagram becomes relevant in two situations, {\it (i)} at 
large $\tan\beta$ and for $m_0 \sim m_{1/2}$, in which case the LSP mass 
is about half the heavy Higgs boson (H and A) mass, and the on-shell 
$s$-channel annihilation is rapid enough to decrease the LSP relic density 
and to allow for larger LSP masses~\cite{funnel}; and {\it (ii)} at any 
large value of  $m_0$, for which there is always a (large) value of 
$\tan\beta$ yielding a small value of $\mu$ through radiative electroweak 
symmetry breaking (focus points~\cite{focus}). 
In the latter configuration, the LSP gets a large Higgsino component, 
hence a large coupling to Higgs bosons. The large value of $\tan\beta$ 
enhances the Higgs couplings to down-type fermion. This conspiracy 
of large couplings yields a large (off-shell) annihilation cross section.

The upper limits of $m_0$ and $m_{1/2}$ can reach 1.5\,TeV/$c^2$ if the 
rapid annihilation is open, and $m_0$ can increase to several TeV/$c^2$
in the parameter sets corresponding to the focus points. 

\subsection{A set of benchmark points}

\begin{figure}[ht]
\begin{picture}(160,80)
\put(-10,-2.5){\epsfxsize85mm\epsfbox{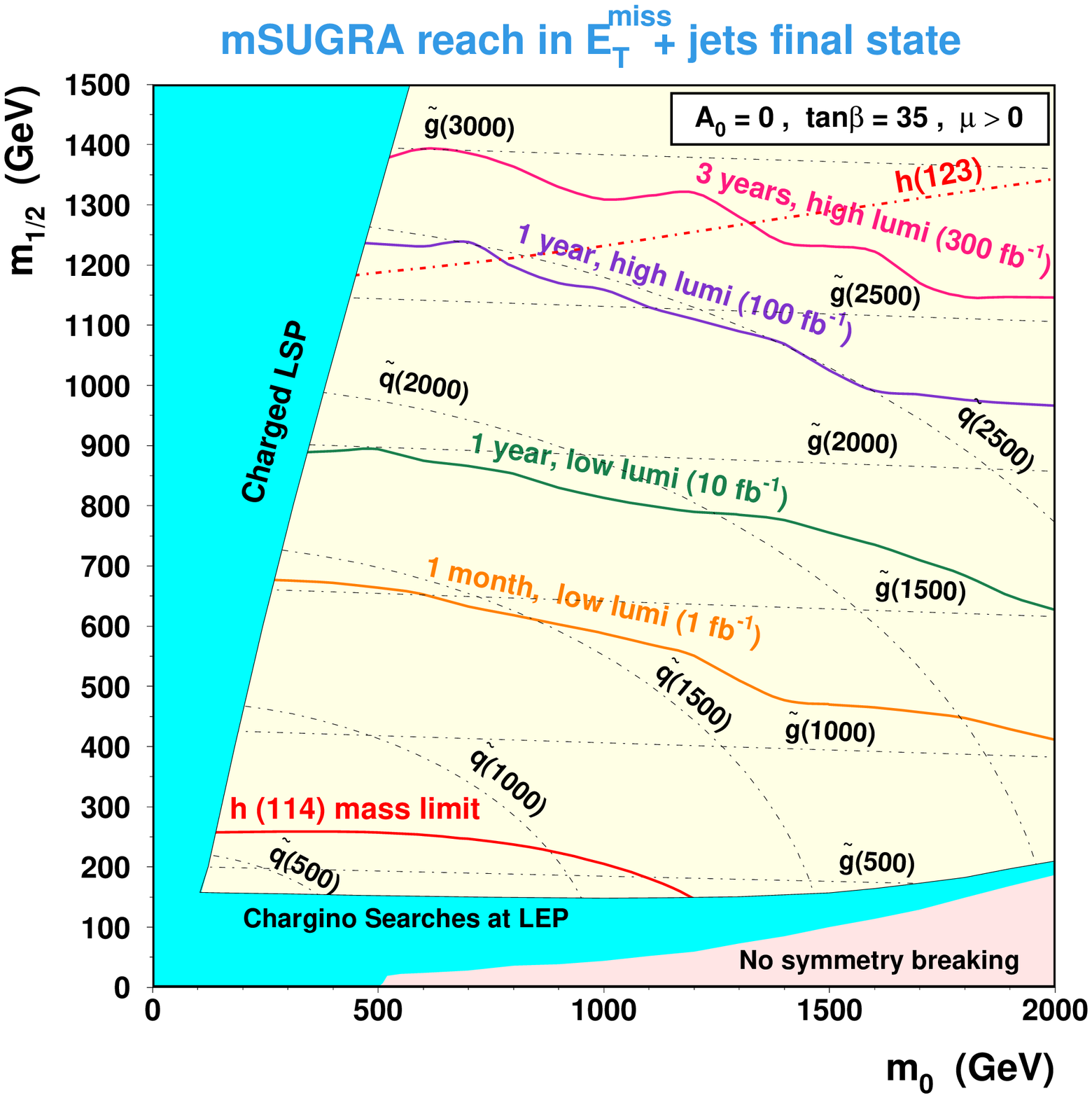}}
\put(-7,-3.5){\epsfxsize87mm\epsfbox{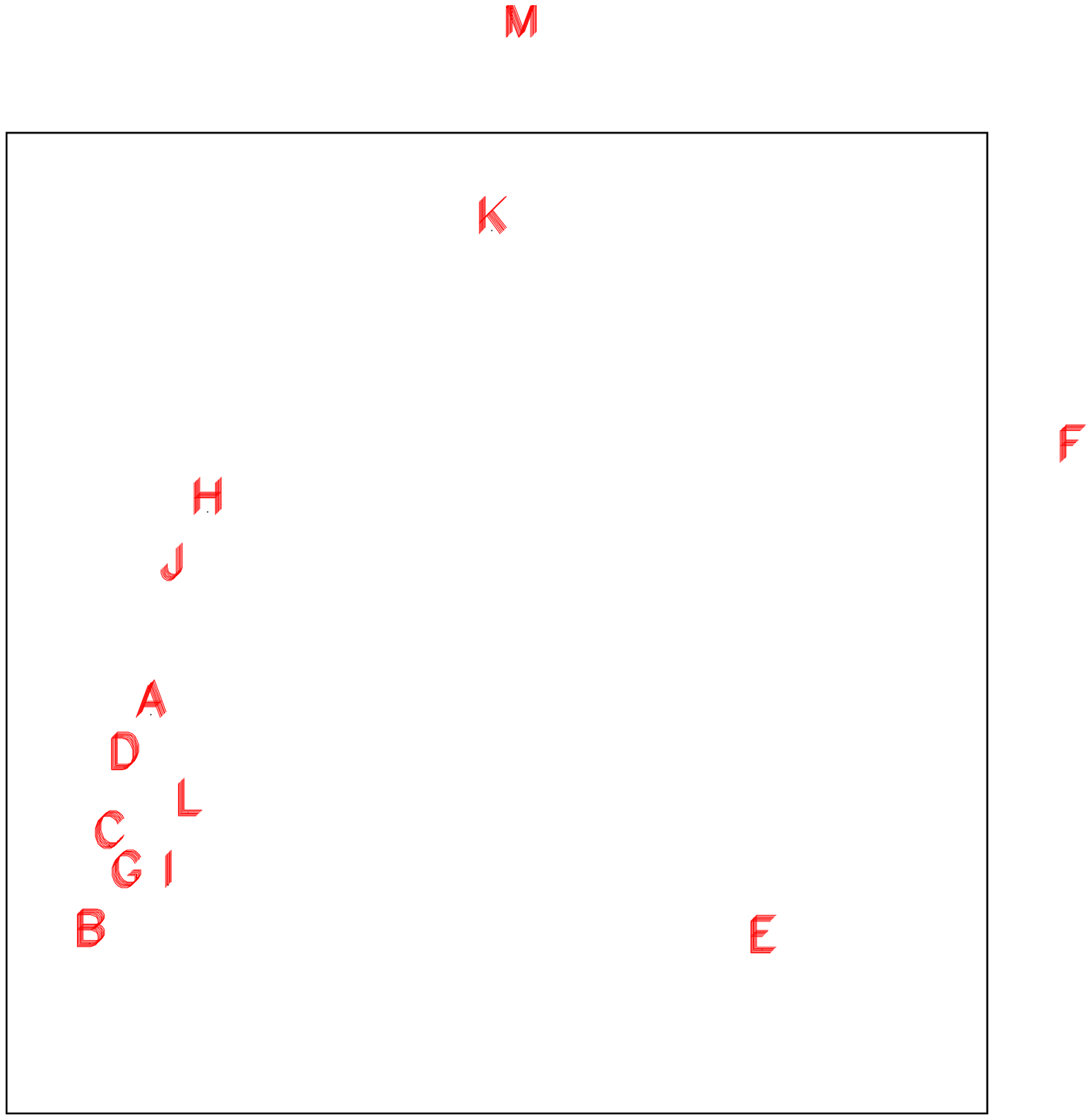}}
\put(80,5){\epsfxsize81mm\epsfbox{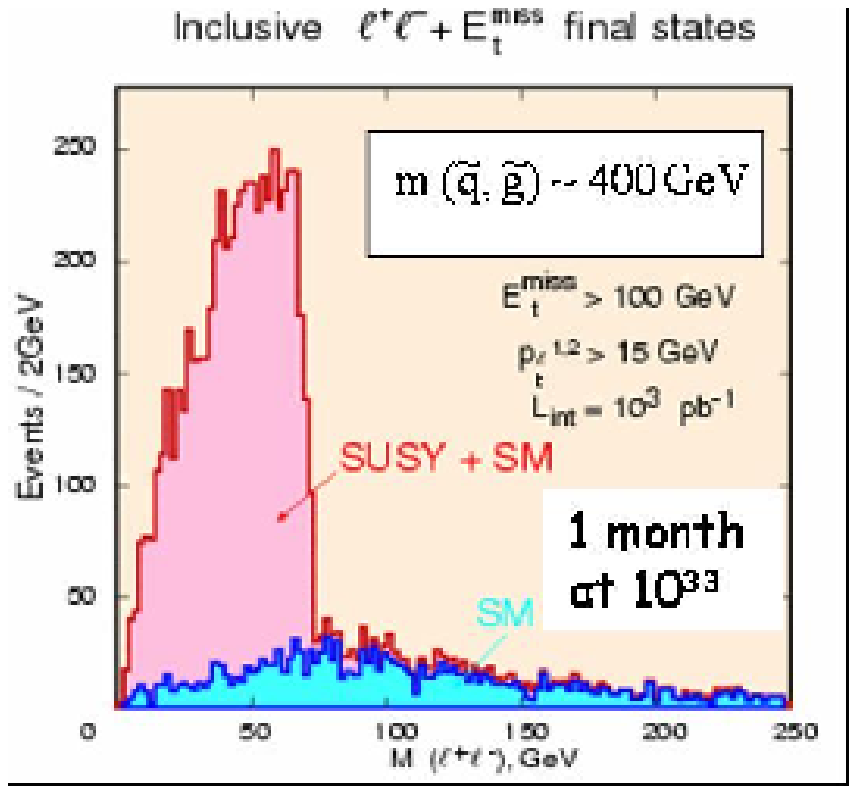}}
\put(5,70){(a)}
\put(91,68){(b)}
\end{picture}
\caption{(a) Benchmark points in the $(m_0,m_{1/2})$ plane and CMS reach with 
searches for final states with jets and missing $E_T$; (b) Dilepton invariant 
mass distribution at point B with 1\,\infb\ in the opposite-sign same-flavour 
final state (light histogram) and in the opposite-sign opposite-flavour 
final state (dark histogram).}
\label{fig:cms}
\end{figure}

The annihilation bulk, the co-annihilation tail, the rapid annihilation 
funnels and the focus-point regions, qualitatively represented in the 
$(m_{1/2},m_0)$ plane in Fig.~\ref{fig:rapid}b, have been populated with 
different sets of benchmark points by various authors. The set of points, 
labelled from A to M and proposed in Ref.~\cite{ellis}, is displayed in 
Fig.~\ref{fig:cms}a on top of the LHC reach~\cite{cmsreach} in the 
$(m_0,m_{1/2})$ plane via the gluino and squark searches in the 
bread-and-butter ``Jets + Missing $E_T$'' final state.
 
It can be seen that the points in the annihilation bulk (A, B, C, D, I, L) 
would lead to a discovery of Supersymmetry with less than 1\,\infb. 
More luminosity may therefore allow detailed measurements to be performed 
there. The co-annihilation tail points (J, H) need up to 10\,\infb\ for 
a discovery, which renders the subsequent measurements more demanding in 
integrated luminosity. Points in the rapid annihilation funnels (K, M) or 
focus points (E, F) require much more luminosity for a discovery, and may 
even not be visible at the LHC (F, M). It should be mentioned that the focus 
points of Ref.~\cite{ellis} were determined with a top quark mass of 
171\,\gevcc, and that $m_0$ and $m_{1/2}$ are rapidly-increasing functions 
of this mass. The recently revised value~\cite{mtop} of $m_{\rm top}$ from 
the TeVatron (about $178 \pm 5$\,\gevcc) expels points E and F well 
beyond the reach of LHC.

\section{A favourable example: Point B}
\label{sec:ptb}

Because it yields the smallest squark and gluino masses, the most favourable 
point for the LHC (as well as for any other collider) in the above set 
is point B. The $\tilde{\rm q}$ and $\tilde{\rm g}$ production cross section 
at this point is so large that it would be visible in the counting room by 
simply watching the Jet-Trigger rate. In addition, as all sparticles
are expected to be light (from 100 to 800\,\gevcc), long cascade
decay chains are possible, e.g., 
\begin{equation}
\tilde{\rm g} \to \tilde{\rm b}\bar{\rm b} \to \chi^0_2 \bbbar
\to \tilde\ell^+ \ell^- \bbbar \to \lplm\bbbar\chi^0_1
\label{eq:chain}
\end{equation}
leading to spectacular final states with b jets, missing 
energy and leptons, which are easy to separate from 
standard-model backgrounds.

\subsection{End-point reconstruction}

In addition to being spectacular, the final states of such a long decay chain
are kinematically constrained by the masses of the numerous sparticles 
involved (here $\chi^0_1$, $\tilde\ell$, $\chi^0_2$, $\tilde{\rm b}$ 
and $\tilde{\rm g}$). For example, the invariant mass of the \lplm\ pair 
(\epem\ or \mpmm) is bounded from above by
\begin{equation}
M_{\lplm}^{\rm max} = { 
\sqrt{
\left( m^2_{\chi^0_2} - m^2_{\tilde\ell} \right)
\left( m^2_{\tilde\ell} - m^2_{\chi^0_1} \right)
}
\over
m_{\tilde\ell}
},
\end{equation}
a value reached in the kinematical configuration in which the lepton from the 
$\tilde\ell \to \ell\chi^0_1$ decay is emitted back-to-back with the 
lepton from the $\chi^0_2 \to \tilde\ell \ell$ decay, in the $\chi^0_2$ rest
frame. As can be seen from Fig.~\ref{fig:cms}b, the distribution of the 
dilepton invariant mass obtained from events with jets, missing $E_T$ and
same-flavour opposite-charge lepton pairs, allows this end point to be 
reconstructed with good accuracy and little 
background~\cite{cmsreach,alessia}. 
This figure corresponds to an integrated luminosity of 1\,\infb, {\it i.e.}, 
about one month running at the design low luminosity value, 
$10^{33}\,{\rm cm}^{-2} {\rm s}^{-1}$.
The latter can be determined, independently of any simulation, from the mass 
distribution obtained with opposite-flavour opposite-charge dilepton 
events.

Other distributions, such as those of the dilepton-jet and the lepton-jet 
invariant masses, do also have end points related to the masses of 
the sparticles involved. For example, the largest dilepton-jet mass among 
all the $\ell\ell$-jet combinations is expected to be bounded from 
above by
\begin{equation}
M_{\ell\ell{\rm q}}^{\rm max} = { 
\sqrt{
\left( m^2_{\chi^0_2} - m^2_{\chi^0_1} \right)
\left( m^2_{\tilde{\rm q}} - m^2_{\chi^0_2} \right)
}
\over
m_{\chi^0_2}
},
\end{equation}
and the largest lepton-jet masses, for the first and the second leptons, 
by
\begin{equation}
M_{\ell_1{\rm q}}^{\rm max} =  { 
\sqrt{
\left( m^2_{\chi^0_2} - m^2_{\tilde\ell} \right)
\left( m^2_{\tilde{\rm q}} - m^2_{\chi^0_2} \right)
}
\over
m_{\chi^0_2}
} \ \ \hbox{and} \ \ \
M_{\ell_2{\rm q}}^{\rm max}  =  { 
\sqrt{
\left( m^2_{\tilde\ell} - m^2_{\chi^0_1} \right)
\left( m^2_{\tilde{\rm q}} - m^2_{\chi^0_2} \right)
}
\over
m_{\tilde\ell} 
}.
\end{equation}
The corresponding distributions can be found in Ref.~\cite{atlasend}, and 
the various end points can be determined with an accuracy of the order 
of 5\,\gevcc\ with an integrated luminosity of 100\,\infb ({\it i.e.},
one full year at the high-luminosity design value, 
$10^{34} {\rm cm}^{-2} {\rm s}^{-1}$). The same kind of measurements, but
with correspondingly reduced statistics, can be repeated with b-tagged 
jets, which give access to mass combinations involving the two sbottom 
masses~\cite{alessia}. 

\subsection{Sparticle mass evaluation}

\begin{figure}[ht]
\begin{picture}(160,85)
\put(0,-15){\epsfxsize100mm\epsfbox{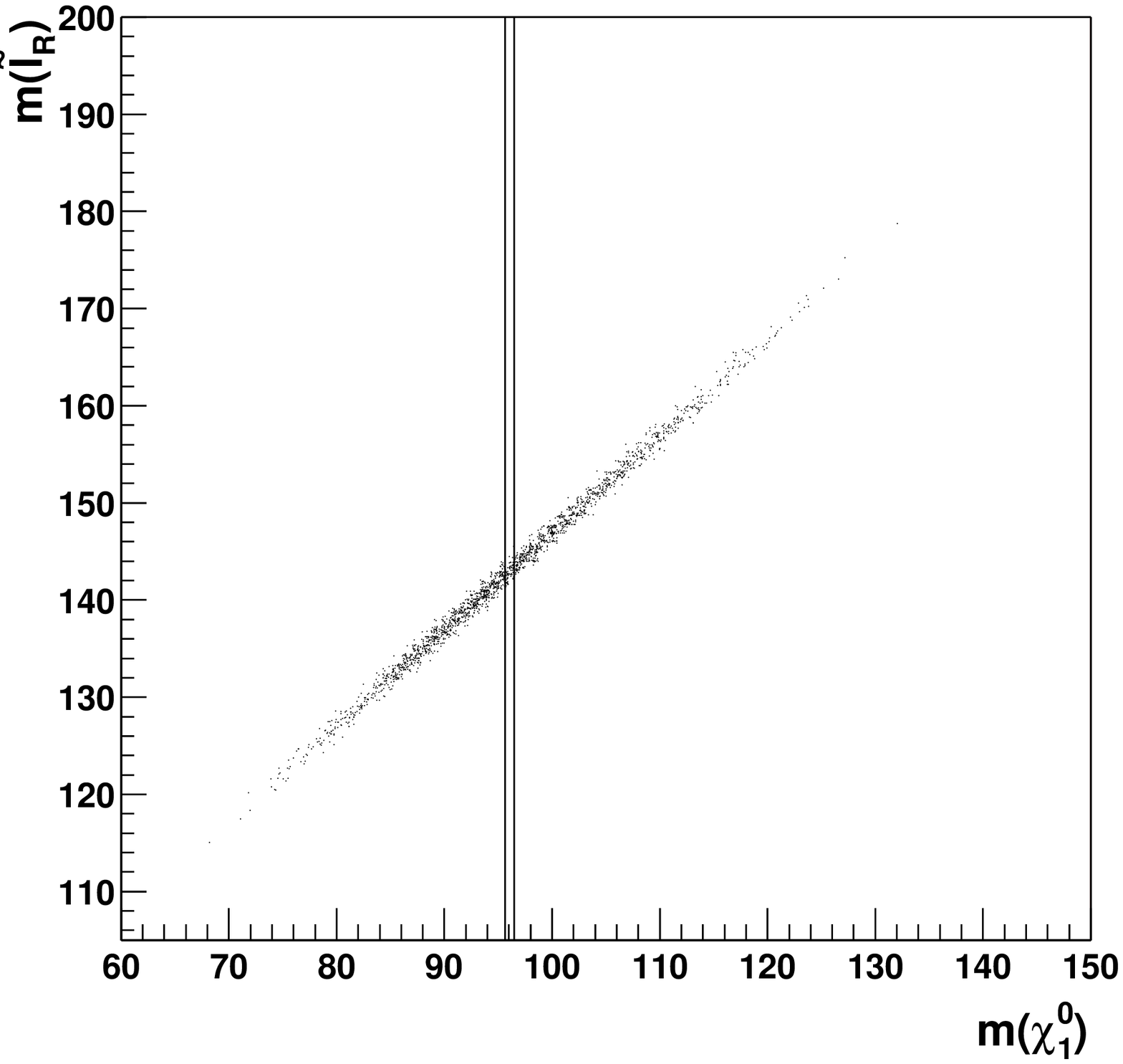}}
\put(91,-1.5){\epsfxsize71mm\epsfbox{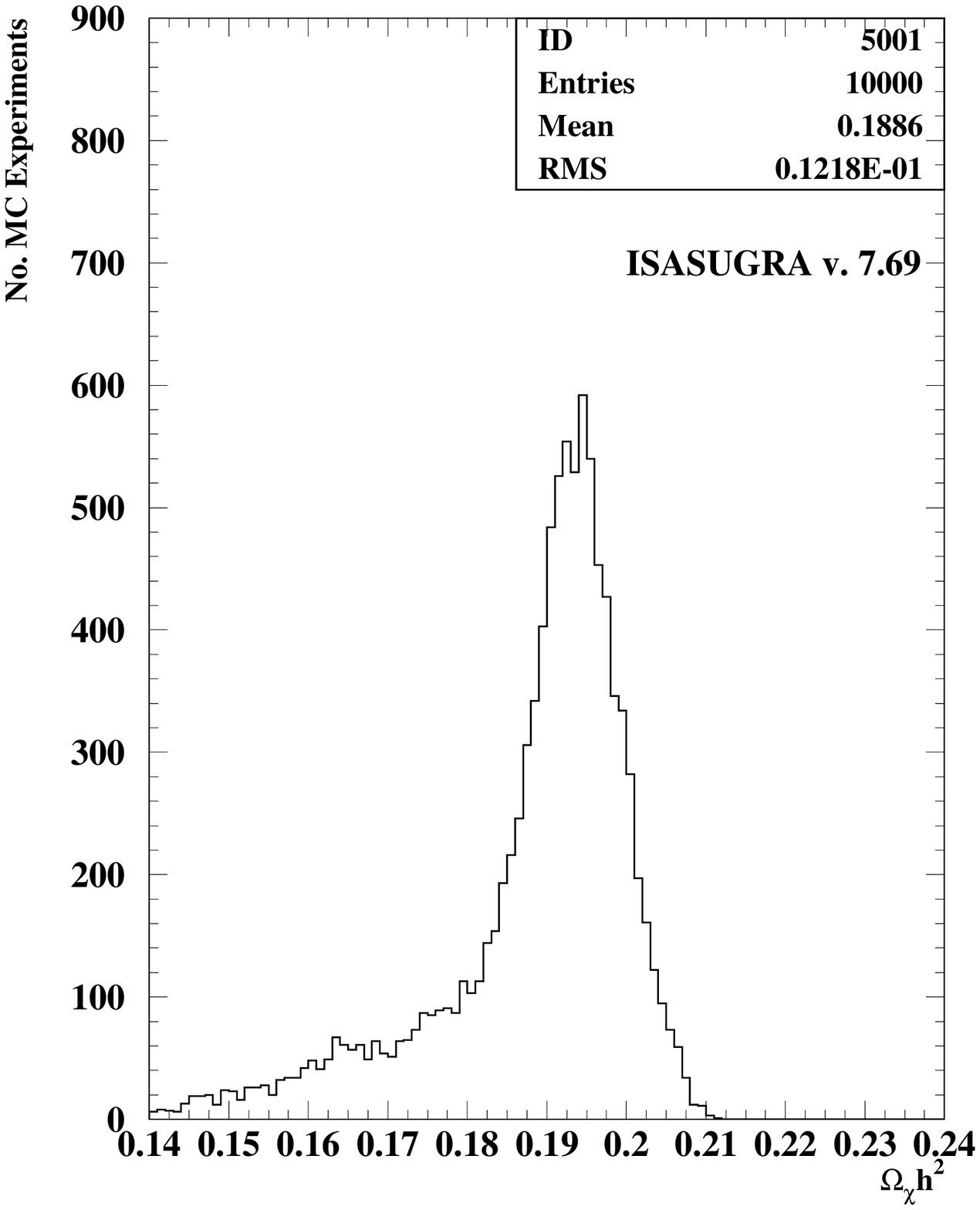}}
\put(12,75){(a)}
\put(106,75){(b)}
\end{picture}
\caption{(a) Correlations between the slepton and the LSP masses 
reconstructed from invariant mass distribution edges, without 
additional theoretical input, at point B, with 300\,\infb\ for each 
point in the graph; (b) Results of the Dark Matter fit 
in mSugra at a point similar to point B, but with an input $\Omega_\chi h^2$ 
value of 0.195, with an integrated luminosity of 100\,\infb\ per entry.}
\label{fig:correl}
\end{figure}

Owing to the length of the decay chain, the number of measurements 
(the distribution edges) turns out to be larger than the number of 
unknowns (the masses of the sparticles involved). It can therefore 
be expected that all sparticle masses can be determined by solving 
an over-constrained system of equations. Because the end-point 
values depend on mass differences rather than on the masses themselves, 
strong correlations between the masses take place~\cite{atlasend}. 
The resolution on the sparticle masses turns out to be worse by a factor 
5 to 10 than that on the edges themselves. An example of such correlations 
is displayed in Fig.~\ref{fig:correl}a for the LSP and the slepton mass 
determination with a large sample of gedanken experiments each representing 
300\,\infb\ of data at the LHC. The vertical line in this figure exemplifies 
the ability of a 350\,GeV \epem\ linear collider to determine the LSP mass 
for the same point B.

To go further with the LHC only, {\it i.e.}, to improve the resolution on 
the sparticle masses and, ultimately, to evaluate the Dark Matter parameters, 
some additional input is needed. What is usually done is to inject some 
theoretical knowledge in the picture, e.g., the assumption that mSugra 
is indeed the proper description of Supersymmetry breaking, and to fit the 
mSugra predictions to the set of measured end-points. The same philosophy 
was adopted, for instance, at LEP to fit the measured Z-lineshape parameters 
to the Standard Model predictions.

\subsection{Dark Matter evaluation in mSugra}

The overall procedure for the evaluation of $\Omega_\chi h^2$ is based on a 
Monte Carlo technique. Many gedanken 
experiments are generated, with all the end-point measurements and their
expected uncertainties, for a specific value of the integrated luminosity, 
{\it e.g.}, 100\,\infb. For each gedanken experiment, 
the best mSugra point is fitted by minimizing the overall $\chi^2$ of 
the end-point measurements. The fitted
parameters allow the LSP mass and the expected LSP relic density to be 
computed, hence the value of $\Omega_\chi h^2$ to be 
determined~\cite{polesello}. The 
distribution of $\Omega_\chi h^2$ for the whole sample of 100\,\infb\ 
gedanken experiments is displayed in Fig.~\ref{fig:correl}b, for a 
parameter set yielding masses and cross sections similar to point B
(but with a relic density of 0.19, i.e., outside the presently allowed 
interval). The long tail at low values disappears with 300\,\infb, when
the two sbottom masses may be disentangled, yielding an experimental 
accuracy of 2.5\% on $\Omega_\chi h^2$, appropriate for a comparison with the 
CMB measurement.

To summarize, in the most favourable point of the annihilation bulk, 
the full LHC luminosity is needed to reach a level of accuracy deemed 
adequate for a sound Dark Matter evaluation, if mSugra is to be 
considered as the theory describing SUSY breaking. 

\subsection{A possible improvement: Full mass reconstruction}
\label{sec:full}

Because only events at the edges of mass distributions are considered,
the end-point technique is affected by an important loss of statistics.
In the decay chain of Eq.~\ref{eq:chain}, each event can be optimally 
used in a set of five equations~\cite{full},
\begin{eqnarray}
m_{\chi^0_1}^2 & = &  p_{\chi^0_1}^2, \\
m_{\tilde\ell}^2 & = & (p_{\chi^0_1} + p_{\ell_1})^2, \\
m_{\chi^0_2}^2 & = & (p_{\chi^0_1} + p_{\ell_1} + p_{\ell_2})^2, \\
m_{\tilde{\rm b}}^2 & = & (p_{\chi^0_1} + p_{\ell_1} + p_{\ell_2}
+p_{{\rm b}_1})^2, \\
m_{\tilde{\rm g}}^2 & = & (p_{\chi^0_1} + p_{\ell_1} + p_{\ell_2}
+p_{{\rm b}_1}+p_{{\rm b}_2})^2,
\end{eqnarray}
for nine unknowns, namely the five sparticle masses, and the LSP four-momentum.
(The four-momenta of the two leptons and the two b jets are measured.) These
five equations therefore yield, in principle, one relation between the five 
masses for each single event. Consequently, all masses can be determined with 
five events only. 

Because all events can be used (not only the edge events), 
the statistics are expected to be a factor of about three higher than 
in the edge technique. This principle, however, has still to be tested with 
real simulations to properly evaluate its potential.

\section{More difficult cases for LHC}
\label{sec:difficult}

The case described in Section 2 is valid for the most favourable point B, for 
which sparticle masses are small and branching ratios into \epem\ and \mpmm\
are large. For this point, the LHC full luminosity is just enough to come to 
a conclusion on Dark Matter, if a sufficient amount of theory is injected 
in the process.

The next-to-most favourable point (point G) already presents more difficulties
because the leptonic branching fraction $\chi^0_2 \to \chi^0_1 \lplm$
is about ten times smaller than at point B. As a consequence, ten times more
statistics ($1\,{\rm ab}^{-1}$) and the full mass reconstruction method
of Section~\ref{sec:full}, are needed to say something relevant about 
Dark Matter. This integrated luminosity is beyond the current expecations 
for the LHC.

The next point (point I) is not an easier case because $\chi^0_2$ decays 
to $\chi^0_1 \tptm$ with a branching ratio close to 100\%. The full mass 
reconstruction is therefore not possible, because the missing neutrinos 
prevent the four-momenta of the taus from being measured. The hope is 
that some end points may still be measurable in the 
$\tau \to {\rm a}_1 \nu_\tau$ decay, 
and that other decay channels, such as $\chi_2^0 \to \chi_1^0 {\rm h} \to 
\chi_1^0 \bbbar$ could be exploited as well. In both cases, more work is 
needed (and is being done) to come to a meaningful conclusion.

Some other points in the co-annihilation tail (D, J, L) have the 
same characteristics as point I. The small mass difference between the 
LSP and the stau, a general feature in the co-annihilation tail, 
does not help, as it tends to produce soft $\tau$'s. The larger sparticle 
masses further enhance the difficulty. Other points (A, C, H) are more 
similar to point B, but larger sparticle masses increase the luminosity 
needed for a sound Dark Matter measurement by factors from 10 to 100, 
{\it i.e.}, well beyond the estimated capabilities of LHC.

As already mentioned, in most of the focus points and in the rapid 
annihilation funnels, the sparticles are too heavy to be seen 
at the LHC, especially with the recently revised top mass value from the 
TeVatron. In these points, only the lighter Higgs boson would be seen 
at the LHC. As pointed out at this conference~\cite{biquette}, a sub-TeV 
\epem\ linear collider would very much help in point B (and in a few 
other points of the annihilation bulk and the co-annihilation tail) with 
an accurate determination of the LSP mass, thus alleviating the need 
of additional theoretical assumptions in the Dark Matter and sparticle 
mass fits.

\section{Conclusions and Outlook}

In the present state of the art, full sparticle reconstruction at LHC 
and confrontation with space measurements are possible in favourable
parameter sets of the the so-called annihilation bulk. An 
integrated luminosity of 100\,\infb\ or more ({\it i.e.} the full design 
LHC luminosity) is needed for a sound comparison of LHC measurements with 
CMB. In the co-annihilation tail, more work is needed to include decays of 
the next-to-lightest neutralino to taus or to Higgs bosons. The 
full mass reconstruction method will undoubtedly help to best exploit the 
LHC-design integrated luminosity of 300\,\infb. The focus points and the rapid 
annihilation funnels are essentially hopeless: there are no long decay 
chains available, and gluinos, squarks and sleptons are out of the 
reach of the LHC.

To conclude with a word of caution, it should be noted that existing studies
are based on mSugra predictions, with R-parity conservation and in general 
use fast detector simulations. If Supersymmetry breaking were to be described
in the framework of a non-constrained MSSM (unlike mSugra), the situation
would become really intricate. (The study of other SUSY breaking mechanisms, 
like Gauge-Mediated SUSY Breaking, is just starting, and no conclusions are 
yet available.) Of course,
other/additional Dark Matter sources would change the picture significantly. 
Finally, the use of full detector simulations will 
certainly give rise to a series of additional systematic uncertainties
(related, {\it e.g.}, to b-tagging performance, missing transverse 
energy resolution, detector alignment and calibration) which are 
yet to be evaluated.

Dark Matter studies at the LHC will therefore be challenging, hence
very interesting. Only data will tell what Dark Matter is made of. The 
data from the LHC will bring precious information, but the final word on the 
topic will probably require information from complementary colliders.

\subsection*{Acknowledgments}

It is a real pleasure to thank J. Tran Thanh Van for his hospitality at 
the Rencontres de Moriond, and for his patience while waiting for my 
contribution to these proceedings. I am grateful to John Ellis, Fabiola 
Gianotti and Luc Pape for their help during the preparation of the talk. 
I am indebted to Nancy Marinelli, Luc Pape and Paris Sphicas for a careful 
reading of the manuscript.


\section*{References}
\begin{thebibliography}{99}
%
\bibitem{dama} A.\,Incicchitti, {\em Dama results}, talk given 
at the XXXIXth Recontres de Moriond (Electroweak Interactions and
Unified Theories, March 2004), in these proceedings. 
%
\bibitem{fiorucci} S.\,Fiorucci, {\em Dark Matter review}, talk given 
at the XXXIXth Recontres de Moriond (Electroweak Interactions and
Unified Theories, March 2004), in these proceedings.
%
\bibitem{wmap} C.L.\,Bennet et al., \Journal{APJS}{148}{2003}{1}.
\bibitem{coann} J.R.\,Ellis, T.\,Falk and K.A.\,Olive, 
\Journal{\PLB}{444}{1998}{367}.
%
\bibitem{funnel} M.\,Drees and M.M.\,Nojiri, 
\Journal{\PRD}{47}{1993}{376}.
%
\bibitem{focus} J.L.\,Feng, K.T.\,Matchev and F.\,Wilczek, 
\Journal{\PLB}{482}{2000}{388}.
%
\bibitem{ellis} M.\,Battaglia et al., \Journal{\EPJC}{33}{2004}{273}.
%
\bibitem{cmsreach} S.\,Abdullin et al. (CMS Coll.), 
\Journal{\JPG}{28}{2002}{469}.
%
\bibitem{mtop} L.\,Cerrito, {\em Top quark mass measurements}, talk given 
at the XXXIXth Recontres de Moriond (QCD and High-Energy Hadronic 
Interactions, April 2004), to appear in the proceedings.
%
%
\bibitem{alessia} M.\,Chiorboli and A.\,Tricomi (CMS Coll.), {\em Squark 
and gluino reconstruction with the CMS detector},
CMS Rapid Note {\bf CMS-RN-2003/002}.
%
\bibitem{atlasend} B.K.\,Gjelsten et al., {\em A detailed analysis of 
the measurement of SUSY masses with the ATLAS detector at the LHC}, 
ATLAS Physics Note {\bf ATL-PHYS-2004-007}.
%
\bibitem{polesello} G.\,Polesello and D.R.\,Tovey, {\em Constraining 
SUSY Dark Matter with the ATLAS detector at the LHC}, ATLAS
Physics Note {\bf ATL-PHYS-2004-008}.
%
\bibitem{full} M.M.\,Nojiri, G.\,Polesello and D.R.\,Tovey, {\em Proposal
for a new reconstruction technique for SUSY processes at the LHC},
ATLAS Physics Note {\bf ATL-PHYS-2003-039}.
%
\bibitem{biquette} G.\,Moortgat-Pick, {\em SUSY parameters in a combined 
LHC+LC analysis}, talk given 
at the XXXIXth Recontres de Moriond (Electroweak Interactions and
Unified Theories, March 2004), in these proceedings. 
%
\end{thebibliography}
\end{document}